\documentclass[aps,prl,twocolumn,showpacs,preprintnumbers,amsmath,amssymb,groupedaddress,longbibliography]{revtex4-2}
\usepackage{hyperref}
\hypersetup{colorlinks=true, citecolor=blue, urlcolor=blue, linkcolor=blue,urlcolor=cyan}
\usepackage{color}
\usepackage{graphicx}
\usepackage{dcolumn}
\usepackage{braket}
\usepackage{amsthm}
\usepackage[caption=false]{subfig}

\def\be{\begin{equation}}
\def\ee{\end{equation}}
\def\ba{\begin{eqnarray}}
\def\ea{\end{eqnarray}}

 \def\ba{{\bar{\alpha}}}

\begin{document}

\title{Stability and Chaotic Dynamics in a Nonlinearly Confined ghost-sector Hamiltonian}

\author{Zahra Molaee}
\affiliation{\it 
School of Astronomy, Institute for Research in Fundamental Sciences (IPM), P.O. Box 19395-5531, Tehran, Iran
}

\newcommand{\SUT}{Department of Physics, Shahid Beheshti University, 1983969411, Tehran, Iran}

\author{Soleyman Fatholahzadeh}
\affiliation{\SUT}

\begin{abstract}

We investigate the nonlinear dynamics of a Hamiltonian system containing a ghost degree of freedom coupled to a canonical sector through nonlinear interactions. The model consists of one positive-energy and one negative-energy mode, augmented by quartic and sextic nonlinearities that regularize the large-amplitude behavior of the system. Unlike conventional ghost models, which typically exhibit runaway trajectories due to the indefinite nature of the kinetic energy, the present Hamiltonian possesses a confining nonlinear structure that renders the accessible energy surfaces compact for finite total energy.

We derive the equations of motion and analyze the geometric properties of the Hamiltonian flow. Particular attention is devoted to the interplay between local ghost-induced instability and global nonlinear confinement. We show that the sextic contribution dominates the asymptotic dynamics and prevents escape to infinity despite the presence of a negative-energy sector. As a consequence, the model provides a controlled framework for studying bounded dynamics in ghost-coupled systems.

The phase-space structure is investigated through numerical integration, Poincaré surfaces of section, and Lyapunov analysis. Depending on the interaction strength and nonlinear couplings, the system exhibits a transition from regular quasi-periodic motion to chaotic dynamics characterized by positive maximal Lyapunov exponents. Remarkably, chaotic trajectories remain confined within compact regions of phase space, yielding a realization of bounded chaotic motion in a ghost-containing Hamiltonian system.

These results provide a concrete example of how nonlinear self-interactions can regularize ghost dynamics at the classical level and indicate a useful prototype for studying stability, confinement, and chaos in nonstandard Hamiltonian theories.

\end{abstract}

\keywords{Ghost dynamics, Hamiltonian systems, nonlinear stability, bounded chaos, Lyapunov exponent, Poincar'e section}

\maketitle

{\bf Introduction.---} 
Hamiltonian systems provide one of the central mathematical frameworks of theoretical physics, describing conservative dynamics in contexts ranging from celestial mechanics to quantum field theory \cite{KAM, Goldstein}. Their phase-space structure often exhibits a rich interplay between regular and chaotic motion, arising from the competition between integrable dynamics and nonlinear perturbations. In near-integrable regimes, the Kolmogorov--Arnold--Moser (KAM) theorem predicts the persistence of a large measure of invariant tori under sufficiently small perturbations \cite{KAM}, whereas stronger nonlinearities may lead to resonance overlap, chaotic transport, and the progressive destruction of regular structures.

A particularly challenging class of Hamiltonian systems involves degrees of freedom with negative kinetic energy, commonly known as ghost modes. Such variables appear in a variety of theoretical settings, including higher-derivative theories, modified gravity, effective field theories, and cosmological models. The conventional concern regarding ghosts originates from the indefinite nature of the corresponding energy functional. Since positive- and negative-energy sectors can exchange energy without bound, ghost-containing systems are often expected to exhibit runaway solutions and severe dynamical instabilities \cite{DeFelice2006, Capozziello2013, Smilga2017, Damour2022}.

Recent studies, however, have suggested that this standard picture is not universally valid. In particular, suitably chosen nonlinear interactions can dramatically modify the dynamics and prevent runaway evolution, even in the presence of an indefinite Hamiltonian structure \cite{vikman, vikman2, BoundedChaos2026}. These findings indicate that nonlinear self-interactions may act as an effective confinement mechanism, restricting trajectories to finite regions of phase space and thereby restoring global boundedness. Such results naturally raise a deeper question: once runaway behavior is suppressed, what is the dynamical structure of the resulting bounded phase space? More specifically, can a ghost-containing Hamiltonian system support complex phenomena such as deterministic chaos while remaining numerically bounded?

The present work addresses this question through the analysis of a nonlinear Hamiltonian model containing one canonical and one ghost degree of freedom. The Hamiltonian combines quadratic terms that generate local instability with higher-order quartic and sextic interactions that dominate at large amplitudes and provide nonlinear confinement. This setting offers a natural framework for investigating the competition between destabilizing ghost dynamics and stabilizing nonlinear effects.

The study builds upon our previous work \cite{BoundedChaos2026}, where we demonstrated that higher-order self-interactions can prevent runaway behavior and generate bounded trajectories in ghost-containing Hamiltonian systems. Here, we move beyond the question of boundedness itself and undertake a detailed investigation of the associated phase-space structure. We derive explicit conditions under which finite-energy manifolds are compact, analyze the equilibrium points and their bifurcations, and characterize the transition from regular to chaotic dynamics using a combination of Poincaré sections, Lyapunov exponents,  frequency drift estimate, and the Mean Exponential Growth factor of Nearby Orbits (MEGNO).

Our central result is the demonstration that bounded chaotic motion can arise in a Hamiltonian system containing a ghost degree of freedom. The existence of such trajectories shows that negative-energy modes do not necessarily imply runaway evolution. Instead, nonlinear interactions can confine the dynamics to compact configuration-space projections  while simultaneously generating a rich phase-space organization that includes both regular and chaotic regions. These findings provide new insight into the nonlinear dynamics of indefinite Hamiltonian systems and broaden the
range of behaviors that ghost-containing models may exhibit.

The paper is organized as follows. Section II introduces the ghost-coupled Hamiltonian model, derives the equations of motion, and analyzes the structure of the potential energy. We determine the equilibrium points, compute the Hessian matrix, and classify the equilibria according to their local stability properties. Section III indicates the conditions for asymptotic confinement and suggests the compactness of the finite-energy shells, leading to global boundedness of trajectories despite the presence of a ghost degree of freedom. Section IV is devoted to linear stability and bifurcation analysis, including the spectral properties of the linearized system, the stability of the equilibrium branches, and the identification of critical parameter values associated with stability-changing bifurcations. Section V investigates the near-integrable regime and the breakdown of regular motion, discussing invariant tori, frequency diffusion, and the onset of chaos from a KAM-theoretic perspective. Section VI presents the numerical methodology and diagnostics used throughout the paper, including phase portraits, Poincaré surfaces of section, Lyapunov exponents, frequency-map analysis, MEGNO indicators, convergence tests, and long-time boundedness verification. Section VII examines the bifurcation structure and the transition from regular to chaotic dynamics through resonance overlap and stochastic transport. Finally, Section VIII summarizes the main results and discusses their implications for the stabilization of ghost dynamics, bounded chaos, and nonlinear confinement in Hamiltonian systems with indefinite kinetic structure.

\section{Hamiltonian Model and Equations of Motion}
In order to investigate the interplay between ghost-induced instability and nonlinear confinement, we consider a Hamiltonian system with two degrees of freedom. The first degree of freedom represents a conventional positive-energy mode, while the second corresponds to a ghost mode characterized by a negative kinetic-energy contribution \cite{DeFelice2006, Damour2022}. The Hamiltonian is defined as

\begin{equation}
H(x,y,p_x,p_y)=
\frac{p_x^2}{2}
-\frac{p_y^2}{2}
+
V(x,y),
\label{eq:Hamiltonian}
\end{equation}

where the potential energy is given by

\begin{equation}
V(x,y)
=
\frac{\omega_x^2}{2}x^2
-\frac{\omega_y^2}{2}y^2
+\lambda x^2 y^2
+\alpha (x^2+y^2)^2
+\eta (x^6+y^6).
\label{eq:Potential}
\end{equation}

where $(x,p_x)$ denotes the canonical sector and $(y,p_y)$ represents the ghost sector. The parameters $\omega_x$ and $\omega_y$ determine the characteristic frequencies of the linearized dynamics, while $\lambda$ controls the nonlinear coupling between the two modes. The quartic coefficient $\alpha$ and sextic coefficient $\eta$ govern the large-amplitude behavior of the system.
The structure of Eq.~(\ref{eq:Potential}) is motivated by the desire to combine local instability with global confinement. The quadratic contribution contains the term

\begin{equation}
-\frac{\omega_y^2}{2}y^2,
\end{equation}

which generates a local instability along the ghost direction. In the absence of higher-order interactions, this term would allow trajectories to escape toward arbitrarily large values of $|y|$. The nonlinear terms are introduced precisely to counteract this tendency at sufficiently large amplitudes.

The quartic interaction
\begin{equation}
\lambda x^2 y^2
\end{equation}

couples the canonical and ghost sectors and provides the primary mechanism through which energy is exchanged between the two modes. Such nonlinear couplings are known to generate resonance phenomena and may lead to the breakdown of regular motion through the destruction of invariant tori.

The quartic self-interaction

\begin{equation}
\alpha (x^2+y^2)^2
\end{equation}

introduces isotropic nonlinear confinement in configuration space. However, the most important contribution to the asymptotic dynamics is the sextic term

\begin{equation}
\eta (x^6+y^6),
\end{equation}

which dominates all lower-order terms for sufficiently large amplitudes.
The Hamiltonian (\ref{eq:Hamiltonian}) is invariant under the discrete transformations

\begin{equation}
x\rightarrow -x,
\qquad
y\rightarrow -y,
\end{equation}

which implies a reflection symmetry with respect to both coordinate axes. This symmetry will be reflected in the structure of the equilibrium points and in the geometry of the phase-space trajectories discussed in subsequent sections.

The model therefore provides a minimal framework in which three competing ingredients coexist:

\begin{enumerate}
\item a local ghost instability generated by the negative quadratic term;
\item nonlinear mode coupling through the interaction $\lambda x^2y^2$;
\item global confinement induced by the positive quartic and sextic nonlinearities.
\end{enumerate}

The competition among these mechanisms gives rise to a rich phase-space structure and creates the possibility of bounded chaotic dynamics\cite{BoundedChaos2026}. 

In the following section we derive the equations of motion and investigate the equilibrium structure of the system.The corresponding Hamilton equations are

\begin{align}
\dot{x} &= p_x, \\
\dot{p}_x &=
-\omega_x^2 x
-2\lambda x y^2
-4\alpha x(x^2+y^2)
-6\eta x^5, \\
\dot{y} &= -p_y, \\
\dot{p}_y &=
\omega_y^2 y
-2\lambda x^2 y 
-4\alpha y(x^2+y^2)
-6\eta y^5.
\end{align}

Combining these equations yields the second-order form

\begin{equation}
\begin{aligned}
\ddot{x} &=
-\omega_x^2 x
-2\lambda x y^2
-4\alpha x(x^2+y^2)
-6\eta x^5, \\
\ddot{y} &=
-\omega_y^2 y
+2\lambda x^2 y
+4\alpha y(x^2+y^2)
+6\eta y^5.
\end{aligned}
\end{equation}

An important feature of the model is its asymptotic behavior. For large amplitudes,

\begin{equation}
V(x,y)
\sim
\eta(x^6+y^6),
\qquad
r\rightarrow\infty.
\end{equation}

Therefore, for $\eta>0$,

\begin{equation}
V(x,y)\rightarrow +\infty,
\qquad
r=\sqrt{x^2+y^2}\rightarrow\infty.
\end{equation}

This property implies that the configuration-space projection of every finite-energy level set remains bounded. Consequently, the nonlinear sextic interaction acts as a global confinement mechanism that suppresses escape to infinity despite the local destabilizing influence of the ghost degree of freedom\cite{Damour2022, BoundedChaos2026}.

The competition between local instability and global confinement constitutes the central dynamical ingredient of the model and forms the basis of the subsequent analysis.

\subsection{Equilibrium Points}

Equilibrium configurations correspond to stationary solutions satisfying

\begin{equation}
\begin{aligned}
\dot{x} &= 0, \\
\dot{y} &= 0, \\
\dot{p}_x &= 0, \\
\dot{p}_y &= 0.
\end{aligned}
\end{equation}
Consequently,

\begin{equation}
p_x=p_y=0,
\end{equation}

while the coordinates must satisfy

\begin{align}
x \Big[
\omega_x^2
+2\lambda y^2
+4\alpha(x^2+y^2)
+6\eta x^4
\Big]
&= 0,
\label{eq:eq1} \\
y \Big[
-\omega_y^2
+2\lambda x^2
+4\alpha(x^2+y^2)
+6\eta y^4
\Big]
&= 0.
\label{eq:eq2}
\end{align}

The equilibrium set naturally decomposes into several branches. Since, each equation is factorized, equilibria arise from different combinations of vanishing coordinates and non vanishing coordinates; therefore, the equilibrium set splits into several branches that can be analyzed separately.

\subsubsection{The Origin}

The trivial equilibrium

\begin{equation}
E_0=(0,0)
\end{equation}

always exists independently of the parameter values.

\subsubsection{Pure Ghost Branch}

Setting $(x=0)$, Eq.~(\ref{eq:eq2}) becomes

\begin{equation}
y
\Big[
-\omega_y^2
+4\alpha y^2
+6\eta y^4
\Big]
=0.
\end{equation}

Besides the origin, nontrivial equilibria satisfy
\begin{equation}
6\eta y^4 + 4\alpha y^2-\omega_y^2
= 0,
\end{equation}

Introducing $(z=y^2)$, one obtains
\begin{equation}
6\eta z^2 + 4\alpha z -\omega_y^2= 0,
\end{equation}

whose positive solution is

\begin{equation}
z_* =
\frac{
-2\alpha +
\sqrt{4\alpha^2 + 6\eta\omega_y^2}
}{6\eta}.
\end{equation}

Therefore, the system possesses the pair
\begin{equation}
E_{\pm} = (0, \pm \sqrt{z_*}).
\end{equation}

\subsubsection{Mixed Equilibria}

For $(x\neq0) $ and $(y\neq0)$, Eqs.~(\ref{eq:eq1})--(\ref{eq:eq2}) reduce to

\begin{align}
\omega_x^2
+2\lambda y^2
+4\alpha(x^2+y^2)
+6\eta x^4
&=0,
\\
-\omega_y^2
+2\lambda x^2
+4\alpha(x^2+y^2)
+6\eta y^4
&=0.
\end{align}

These equations define additional equilibrium branches whenever the parameter values permit simultaneous solutions. In general they must be determined numerically.

\subsection{Hessian Matrix}

The local structure of the potential is governed by the Hessian matrix
\begin{equation}
\mathcal{H}=
\begin{pmatrix}
V_{xx} & V_{xy} \\
V_{yx} & V_{yy}
\end{pmatrix},
\qquad
V_{xy}=V_{yx},
\end{equation}
where
\begin{align}
V_{xx}
&=
\omega_x^2
+2\lambda y^2
+12\alpha x^2
+4\alpha y^2
+30\eta x^4, \\
V_{yy}
&=
-\omega_y^2
+2\lambda x^2
+4\alpha x^2
+12\alpha y^2
+30\eta y^4, \\
V_{xy}
&=
4(\lambda+2\alpha)xy.
\end{align}
\subsection{Classification of Equilibria}

The local character of each equilibrium point is determined by the Hessian matrix of the potential,
\begin{equation}
\mathcal{H}(x,y)=
\begin{pmatrix}
V_{xx} & V_{xy} \\
V_{yx} & V_{yy}
\end{pmatrix},
\qquad
V_{xy}=V_{yx}.
\end{equation}
For a given equilibrium, the signs of the eigenvalues of \(\mathcal{H}\) determine whether the critical point is a local minimum, a local maximum, or a saddle point of the potential.

\subsubsection{The Origin}

At the origin \((0,0)\), the Hessian becomes
\begin{equation}
\mathcal{H}(0,0)=
\begin{pmatrix}
\omega_x^2 & 0 \\
0 & -\omega_y^2
\end{pmatrix}.
\end{equation}
Its determinant is
\begin{equation}
\det \mathcal{H}(0,0)=-\omega_x^2\omega_y^2<0,
\end{equation}
which implies that the Hessian is indefinite. Therefore, the origin is a saddle point of the potential. This is consistent with the presence of a local instability along the ghost direction.

\subsubsection{Pure Ghost Equilibria}

For the pure ghost equilibria \(E_\pm=(0,\pm \sqrt{z_*})\), the mixed derivative vanishes,
\begin{equation}
V_{xy}(E_\pm)=0.
\end{equation}
Hence, the local classification is determined by the diagonal entries
\begin{equation}
V_{xx}(E_\pm), \qquad V_{yy}(E_\pm).
\end{equation}
At these points, one finds
\begin{equation}
V_{xx}(E_\pm)=\omega_x^2+(2 \lambda+4\alpha) z_*,
\end{equation}
and
\begin{equation}
V_{yy}(E_\pm)= -\omega_y^2+12\alpha z_*+30\eta z_*^2,
\end{equation}
where \(z_*\) is the positive root obtained from the equilibrium condition.

For sufficiently large positive \(\alpha\) and \(\eta\), the nonlinear terms may overcome the destabilizing quadratic contribution in the ghost direction, so that \(V_{yy}(E_\pm)>0\). In that case, the equilibria \(E_\pm\) become locally confining along the ghost direction. Their full classification, however, still depends on the sign of \(V_{xx}(E_\pm)\): if both diagonal entries are positive, the equilibrium is a local minimum; if they have opposite signs, the equilibrium is a saddle point.

\subsection{Proposition 1}

\textbf{Proposition.}
Assume

\begin{equation}
\omega_x>0,
\qquad
\omega_y>0,
\qquad
\alpha>0,
\qquad
\eta>0.
\end{equation}

Then:

\begin{enumerate}
\item The origin is always a saddle equilibrium.
\item The ghost branch contains two symmetric nontrivial equilibria $(E_\pm)$.
\item Any additional equilibria occur in symmetry-related pairs generated by the transformations $(x\rightarrow -x)$ and $(y\rightarrow -y)$.
\item The local ghost instability is confined to a neighborhood of the origin and does not determine the asymptotic structure of the potential.
\end{enumerate}

\textit{Proof.}
The first statement follows from the negative determinant of the Hessian at the origin. The second follows from the positive root of the quartic equation defining the pure ghost branch. The third is a consequence of the discrete reflection symmetries of the Hamiltonian. The fourth follows from the fact that the sextic term dominates asymptotically and eventually overwhelms the negative quadratic contribution. \hfill $\square$

\section{Compact Configuration-space Projections and Global Boundedness}

A key question in Hamiltonian systems with a ghost degree of freedom is whether trajectories with finite energy can escape to arbitrarily large amplitudes. In many ghost models, the indefinite energy structure permits runaway transfer between positive- and negative-energy sectors. In the present model, we show that the higher-order nonlinear terms prevent such escape and enforce global confinement in configuration space.

\subsection{Asymptotic Growth of the Potential}
To study  behavior  of the potential \ref{eq:Potential} for large amplitudes, introduce
\[
r^2=x^2+y^2.
\]
Since
\begin{equation}
x^6+y^6 \ge \frac{1}{4}(x^2+y^2)^3 = \frac{1}{4}r^6,
\label{ineq1}
\end{equation}
the sextic term satisfies
\[
\eta(x^6+y^6)\ge \frac{\eta}{4}r^6
\qquad \text{for } \eta>0.
\]
Moreover, the remaining terms can be bounded from below by lower-order powers of \(r\), so there exist constants \(C_1>0\) and \(C_2>0\) such that
\begin{equation}
V(x,y)\ge -C_1 r^2 - C_2 r^4 + \frac{\eta}{4}r^6.
\end{equation}
Because the sextic term dominates the lower-order contributions as \(r\to\infty\), it follows that
\begin{equation}
V(x,y)\to +\infty,
\qquad r\to\infty,
\end{equation}
whenever
\begin{equation}
\eta>0.
\end{equation}

Therefore, the potential is coercive. In particular, for any finite energy \(E\), the configuration-space set
\begin{equation}
\Omega_E
=
\left\{
(x,y)\in\mathbb{R}^2 : V(x,y)\le E
\right\}.
\end{equation}
is bounded. This implies that finite-energy trajectories cannot escape to infinity in configuration space, and the energy shells are compact in the \((x,y)\)-projection.

\textbf{Proposition 2.}
\emph{If \(\eta>0\), then \(\Omega_E\) is compact for every finite \(E\).}

\medskip

\textit{Proof.}
Since \(V\) is continuous and satisfies
\begin{equation}
V(x,y)\to +\infty
\qquad \text{as} \qquad r=\sqrt{x^2+y^2}\to\infty,
\label{asymptoticV}
\end{equation}
there exists, for each finite \(E\), a radius \(R(E)>0\) such that
\begin{equation}
V(x,y)>E
\qquad \text{whenever} \qquad r>R(E).
\end{equation}
Therefore,
\begin{equation}
\Omega_E \subset \overline{B}_{R(E)},
\end{equation}
where \(\overline{B}_{R(E)}\) denotes the closed ball of radius \(R(E)\) in \(\mathbb{R}^2\).

Moreover, since \(V\) is continuous, the set \(\Omega_E = V^{-1}((-\infty,E])\) is closed as the preimage of a closed set. Hence \(\Omega_E\) is both closed and bounded. By the Heine--Borel theorem, \(\Omega_E\) is compact.
\hfill \(\square\)

\subsection{Compactness of the Energy Shells}

For a fixed energy level \(E\), the Hamiltonian energy surface is defined by
\begin{equation}
\Sigma_E=
\left\{
(x,y,p_x,p_y)\in\mathbb{R}^4 :
H(x,y,p_x,p_y)=E
\right\}.
\end{equation}
Since 
\begin{equation}
H(x,y,p_x,p_y)=\frac{p_x^2}{2}-\frac{p_y^2}{2}+V(x,y),
\end{equation}
the energy constraint can be written as
\begin{equation}
p_x^2-p_y^2=2\bigl(E-V(x,y)\bigr).
\label{energy_constraint}
\end{equation}

From the previous subsection, the configuration-space sublevel set   \(\Omega_E\)
is compact when \(\eta>0\). Hence, on \(\Omega_E\), the potential is bounded below, so there exists a constant \(V_{\min}\) such that
\begin{equation}
V_{\min}\le V(x,y)\le E.
\end{equation}
Therefore,
\begin{equation}
p_x^2-p_y^2 \le 2(E-V_{\min}).
\end{equation}

To obtain boundedness of the momenta on a finite-energy trajectory, one uses the fact that the configuration variables remain confined to the compact set \(\Omega_E\), so \(V(x,y)\) is bounded there, and the energy identity then controls the difference \(p_x^2-p_y^2\). In addition, because the kinetic terms enter the Hamiltonian quadratically and the energy is fixed, neither momentum can diverge without forcing the other to diverge in a way incompatible with the same finite value of \(H\). Consequently, both \(p_x\) and \(p_y\) remain bounded on every finite-energy trajectory.

This shows that the full energy shell \(\Sigma_E\) is bounded. Since \(\Sigma_E\) is also closed as the preimage of the closed set \(\{E\}\) under the continuous Hamiltonian \(H\), it follows that \(\Sigma_E\) is compact by the Heine--Borel theorem \cite{Raffi}.

\vspace{2mm}

\textbf{Theorem 1 (compact configuration-space projections ).}

\emph{
Let
\(\omega_x > 0\), \(\omega_y > 0\), \(\alpha > 0\), and \(\eta > 0\).
Then every finite-energy hypersurface
\(\Sigma_E\)
of the Hamiltonian system
\eqref{eq:Hamiltonian}
is compact.
}

\medskip

\textit{Proof.}

The configuration variables are confined to the compact set
$(\Omega_E)$ by Proposition 2.

The Hamiltonian constraint then bounds both momenta.

Hence
$ [
(x,y,p_x,p_y)
] $
belongs to a closed and bounded subset of
$(\mathbb{R}^4)$.

By the Heine--Borel theorem,
$(\Sigma_E)$ is compact.
\hfill $\square$

From a dynamical-systems perspective, the quadratic part of the Hamiltonian describes two oscillatory modes with opposite energy signatures, while the quartic and sextic interactions deform the underlying harmonic structure and generate nonlinear confinement. In this sense, the model defines a class of confined ghost-coupled harmonic oscillators for which the finite-energy phase-space manifolds remain compact despite the presence of an indefinite kinetic sector. This interpretation places the system within the broader family of nonlinear oscillator models while highlighting the distinctive role played by the ghost mode and the confining higher-order interactions.

\section{Linear Stability and Bifurcation Analysis}

The boundedness results obtained in the previous section guarantee that trajectories cannot escape to infinity for finite energy. Nevertheless, boundedness alone does not determine the qualitative nature of the dynamics. In order to understand the origin of regular and chaotic motion, it is necessary to analyze the local stability properties of the equilibrium points and their evolution under parameter variations.

\subsection{Linearization Around Equilibrium Points}

Let
\begin{equation}
Z = (x,y,p_x,p_y)^T
\end{equation}
denote a phase-space vector and let
\begin{equation}
Z_* = (x_*,y_*,0,0)
\end{equation}
be an equilibrium configuration.

Introducing a small perturbation
\begin{equation}
Z(t) = Z_* + \delta Z(t),
\end{equation}
the equations of motion can be linearized as
\begin{equation}
\dot{\delta Z} = J(Z_*)\,\delta Z,
\end{equation}
where \(J(Z_*)\) is the Jacobian matrix evaluated at the equilibrium point.

Using Hamilton's equations for
\begin{equation}
H(x,y,p_x,p_y)=\frac{p_x^2}{2}-\frac{p_y^2}{2}+V(x,y),
\end{equation}
one obtains
\begin{equation}
J(Z_*)=
\begin{pmatrix}
0 & 0 & 1 & 0 \\
0 & 0 & 0 & -1 \\
- V_{xx}(x_*,y_*) & - V_{xy}(x_*,y_*) & 0 & 0 \\
V_{xy}(x_*,y_*) & V_{yy}(x_*,y_*) & 0 & 0
\end{pmatrix},
\label{Jacobian}
\end{equation}
where
\begin{align}
V_{xx} &= \omega_x^2 + 2\lambda y^2 + 12\alpha x^2 + 4\alpha y^2 + 30\eta x^4, \\
V_{yy} &= -\omega_y^2 + 2\lambda x^2 + 4\alpha x^2 + 12\alpha y^2 + 30\eta y^4, \\
V_{xy} &= 4(\lambda+2\alpha)xy,
\end{align}
are the components of the Hessian matrix of the potential,
\begin{equation}
\mathcal{H}(x,y)=
\begin{pmatrix}
V_{xx} & V_{xy} \\
V_{xy} & V_{yy}
\end{pmatrix}.
\end{equation}

\subsection{Stability of the Origin}

At the origin,
\begin{equation}
(x_*,y_*)=(0,0),
\end{equation}
the Hessian reduces to
\begin{equation}
V_{xx}(0,0)=\omega_x^2,\quad
V_{yy}(0,0)=-\omega_y^2,\quad
V_{xy}(0,0)=0.
\end{equation}

The Jacobian becomes
\begin{equation}
J_0=
\begin{pmatrix}
0 & 0 & 1 & 0 \\
0 & 0 & 0 & -1 \\
-\omega_x^2 & 0 & 0 & 0 \\
0 & -\omega_y^2 & 0 & 0
\end{pmatrix}.
\end{equation}

The characteristic polynomial factorizes as
\begin{equation}
(\mu^2+\omega_x^2)(\mu^2-\omega_y^2)=0.
\end{equation}
Thus, the eigenvalues are
\begin{equation}
\mu=\pm i\omega_x, \qquad \mu=\pm \omega_y.
\end{equation}
The equilibrium at the origin therefore possesses one elliptic pair and one hyperbolic pair of eigenvalues, and is a center--saddle equilibrium.

\subsection{Stability of the Ghost Equilibria}

Consider the nontrivial equilibria on the pure ghost branch
\begin{equation}
E_\pm = (0,\pm\sqrt{z_*}),
\end{equation}
where \(z_*>0\) satisfies the equilibrium condition
\begin{equation}
-\,\omega_y^2 + 4\alpha z_* + 6\eta z_*^2 = 0.
\end{equation}
Solving this quadratic equation for \(z_*=y_*^2\) yields
\begin{equation}
z_* =
\frac{-4\alpha + \sqrt{16\alpha^2+24\eta\omega_y^2}}{12\eta},
\qquad (\alpha>0,\ \eta>0,\ \omega_y>0).
\end{equation}

At these points one has \(x_*=0\), so
\begin{equation}
V_{xy}(E_\pm)=0,
\end{equation}
and the Jacobian decouples into two independent planar subsystems.

Using the general expressions for the Hessian, one finds
\begin{align}
V_{xx}(E_\pm) &= \omega_x^2 + (2\lambda+4\alpha)\,z_*, \\
V_{yy}(E_\pm) &= -\omega_y^2 + 12\alpha\,z_* + 30\eta\,z_*^2.
\end{align}
In particular, \(V_{yy}(E_\pm)\) is strictly positive for sufficiently large \(\alpha>0\) and \(\eta>0\), overcoming the destabilizing quadratic term in the ghost direction.

The linear stability of \(E_\pm\) is determined by the signs of \(V_{xx}(E_\pm)\) and \(V_{yy}(E_\pm)\). When both are positive, the eigenvalues in each sector form an elliptic pair, and the equilibria are linearly elliptic.

A convenient way to express the condition on \(\lambda\) is to introduce the critical value
\begin{equation}
\lambda_c =
-\frac{\omega_x^2+4\alpha z_*}{2 z_*},
\end{equation}
so that
\begin{equation}
V_{xx}(E_\pm)>0
\quad\Longleftrightarrow\quad
\lambda>\lambda_c.
\end{equation}
Provided the parameters \(\alpha,\eta,\omega_y\) are chosen such that \(V_{yy}(E_\pm)>0\), the equilibria \(E_\pm\) are elliptic whenever \(\lambda>\lambda_c\).

\subsection{Bifurcation Structure}

The quantity
\begin{equation}
V_{xx}(E_\pm) = \omega_x^2 + (2\lambda+4\alpha)\,z_*
\end{equation}
changes sign at \(\lambda=\lambda_c\), where
\begin{equation}
\lambda_c =
-\frac{\omega_x^2+4\alpha z_*}{2 z_*}.
\end{equation}
For \(\lambda>\lambda_c\), and for parameter values satisfying \(V_{yy}(E_\pm)>0\), the ghost equilibria remain elliptic. When \(\lambda<\lambda_c\), the sign of \(V_{xx}(E_\pm)\) changes and a loss of elliptic stability occurs, signaling a stability-changing bifurcation in the ghost sector.

\medskip
\noindent
\textbf{Proposition 3.}

Assume
\[
\omega_x>0,\qquad \omega_y>0,\qquad \alpha>0,\qquad \eta>0,
\]
and let \(z_*\) denote the positive root of
\begin{equation}
-\,\omega_y^2 + 4\alpha z_* + 6\eta z_*^2 = 0.
\end{equation}
Define
\begin{equation}
\lambda_c =
-\frac{\omega_x^2+4\alpha z_*}{2 z_*}.
\end{equation}
Then:
\begin{enumerate}
\item The origin is always a center--saddle equilibrium.
\item The ghost equilibria \(E_\pm\) exist and are elliptic whenever \(\lambda>\lambda_c\) and \(V_{yy}(E_\pm)>0\).
\item A stability-changing bifurcation in the ghost sector occurs at \(\lambda=\lambda_c\).
\item In the nonlinear regime, chaotic transport is expected in the vicinity of this threshold due to the interaction between elliptic and hyperbolic structures.
\end{enumerate}

\hfill $\square$

\section{Near-Integrable Dynamics and KAM Breakdown}

The local stability analysis performed in the previous section reveals the existence of elliptic equilibrium points surrounded by bounded oscillatory motion. In Hamiltonian systems, such structures often signal the presence of invariant tori that organize the surrounding phase-space dynamics.

The purpose of this section is to investigate how these regular structures are progressively destroyed as the nonlinear coupling strength increases, ultimately leading to chaotic transport within the bounded phase-space region indicated by Theorem~1.

\subsection{Near-Integrable Limit}

To identify the origin of regular motion, it is useful to consider the weak-coupling regime

\begin{equation}
|\lambda| \ll 1,
\qquad
\alpha \ll 1,
\qquad
\eta \ll 1.
\end{equation}

In this limit the Hamiltonian can be decomposed as

\begin{equation}
H = H_0 + H_1 ,
\end{equation}

where

\begin{equation}
H_0
=
\frac{p_x^2}{2}
-
\frac{p_y^2}{2}
+
\frac{\omega_x^2}{2}x^2
-
\frac{\omega_y^2}{2}y^2
\end{equation}

is the unperturbed Hamiltonian and

\begin{equation}
H_1
=
\lambda x^2y^2
+
\alpha(x^2+y^2)^2
+
\eta(x^6+y^6)
\end{equation}

acts as a nonlinear perturbation.

For sufficiently weak nonlinear coupling, numerical simulations suggest the presence of extended quasiperiodic regions in phase space, consistent with the persistence of regular motion. 
Although the indefinite kinetic structure prevents a direct application of the classical KAM theorem, the dynamics near stable elliptic islands remains qualitatively similar to that of conventional near-integrable Hamiltonian systems.

In this regime the motion is predominantly regular and trajectories exhibit negligible sensitivity to initial conditions.

\subsection{Fundamental Frequencies from the Hamiltonian Flow}

For regular bounded motion, the trajectory generated by the Hamiltonian flow is expected to lie on an invariant torus and to exhibit quasiperiodic time dependence. In this case, the phase-space variables can be represented approximately as Fourier series of the form
\begin{equation}
x(t)=\sum_{\mathbf{k}} A_{\mathbf{k}} e^{i\,\mathbf{k}\cdot \boldsymbol{\Omega} t},
\qquad
y(t)=\sum_{\mathbf{k}} B_{\mathbf{k}} e^{i\,\mathbf{k}\cdot \boldsymbol{\Omega} t},
\end{equation}
where \(\boldsymbol{\Omega}=(\Omega_x,\Omega_y)\) denotes the vector of fundamental frequencies associated with the orbit.

These frequencies are not parameters of the Hamiltonian. Rather, they are dynamical quantities determined by the solution of the equations of motion. In the weakly nonlinear regime, they vary smoothly with the energy and the initial condition, while in chaotic regions they become time-dependent and exhibit slow diffusion.

In practice, the dominant frequency is extracted from the Fourier spectrum of \(x(t)\) or \(y(t)\). The strongest spectral peak is identified as the principal frequency \(\Omega\). This quantity will be used in the  frequency drift estimate to quantify frequency drift and to distinguish regular from chaotic motion.

\section{Numerical Methods}

To quantify the phase-space structure of the Hamiltonian system, we integrate the equations of motion numerically and compute two standard diagnostics: the relative energy error and the Poincar\'e surface of section. The chosen parameter set produces a bounded orbit in the weakly nonlinear regime, making it suitable for examining both numerical accuracy and the geometry of the trajectory.

\subsection{Integration of the Equations of Motion}

The Hamiltonian \eqref{eq:Hamiltonian} defines the four-dimensional dynamical system
\begin{equation}
\dot{\mathbf Z}=\mathbf F(\mathbf Z), 
\qquad
\mathbf Z=(x,y,p_x,p_y)^T.
\end{equation}
For the numerical experiment reported here, the parameters are fixed as
\begin{equation}
\omega_x=1.0, \omega_y=1.2, \lambda=0.15, \alpha=0.08, \eta=0.02.
\end{equation}
The system is integrated over the interval
\begin{equation}
0\le t\le T_{\max},\qquad T_{\max}=5000,
\end{equation}
using an adaptive Dormand--Prince Runge--Kutta method with
\begin{equation}
\mathrm{RelTol}=\mathrm{AbsTol}=10^{-12}.
\end{equation}
A uniform output grid of \(50001\) points is used for post-processing.

The initial condition is chosen as
\begin{equation}
(x_0,y_0,p_{x0},p_{y0})=(0.10,\,0.05,\,0,\,0),
\end{equation}
which corresponds to the initial energy
\begin{equation}
E_0=0.0032162703125000005.
\end{equation}

\subsection{Energy Conservation}

\begin{figure}[htbp]
    \centering
    \includegraphics[width=0.4\textwidth]{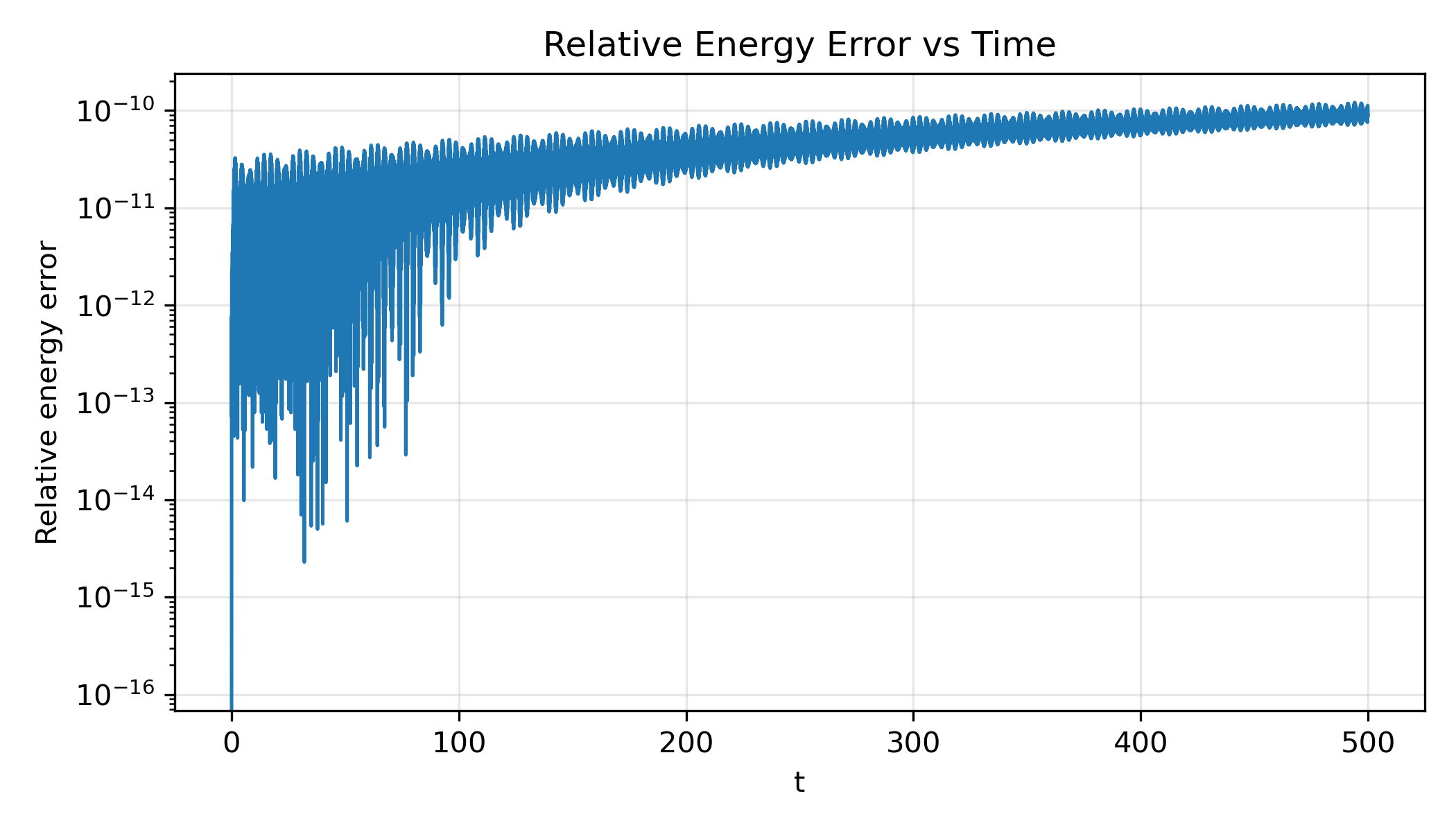}
    \caption{Relative energy error \(\Delta_E(t)\) as a function of time for the orbit with initial condition \((0.10,0.05,0,0)\). The maximum deviation remains at the level of \(10^{-10}\), indicating excellent numerical conservation of the Hamiltonian.}
    \label{fig:energy_error}
\end{figure}

Since the exact flow is Hamiltonian, the energy should remain constant along each trajectory. To monitor the numerical accuracy, we compute the relative energy error
\begin{equation}
\Delta_E(t)=\frac{|H(t)-H(0)|}{|H(0)|}.
\end{equation}
For the trajectory considered here, the maximum relative error is
\begin{equation}
\max_t \Delta_E(t)=1.2102023346109034\times 10^{-10}.
\end{equation}
This confirms that the numerical solution remains very close to the exact energy manifold throughout the integration interval. Figure~\ref{fig:energy_error} shows the time dependence of the relative energy error.

\subsection{Poincaré  Surface of Section}

To visualize the phase-space structure, we construct a Poincar\'e surface \cite{Carone2009b} of section defined by
\begin{equation}
y=0,\qquad p_y>0.
\end{equation}
Whenever the trajectory crosses this hypersurface, the corresponding \((x,p_x)\) values are recorded by linear interpolation between successive integration points. For the present orbit, this procedure yields \(1000\) section points over the interval \(0\le t\le 5000\).

The resulting Poincar\'e section provides a compact representation of the dynamics in the reduced phase space. In the present case, the crossings remain confined to a bounded region, reflecting the bounded nature of the trajectory. The detailed arrangement of the points can be used to distinguish between regular, resonant, and chaotic motion.

\begin{figure}[htbp]
    \centering
    \includegraphics[width=0.46\textwidth]{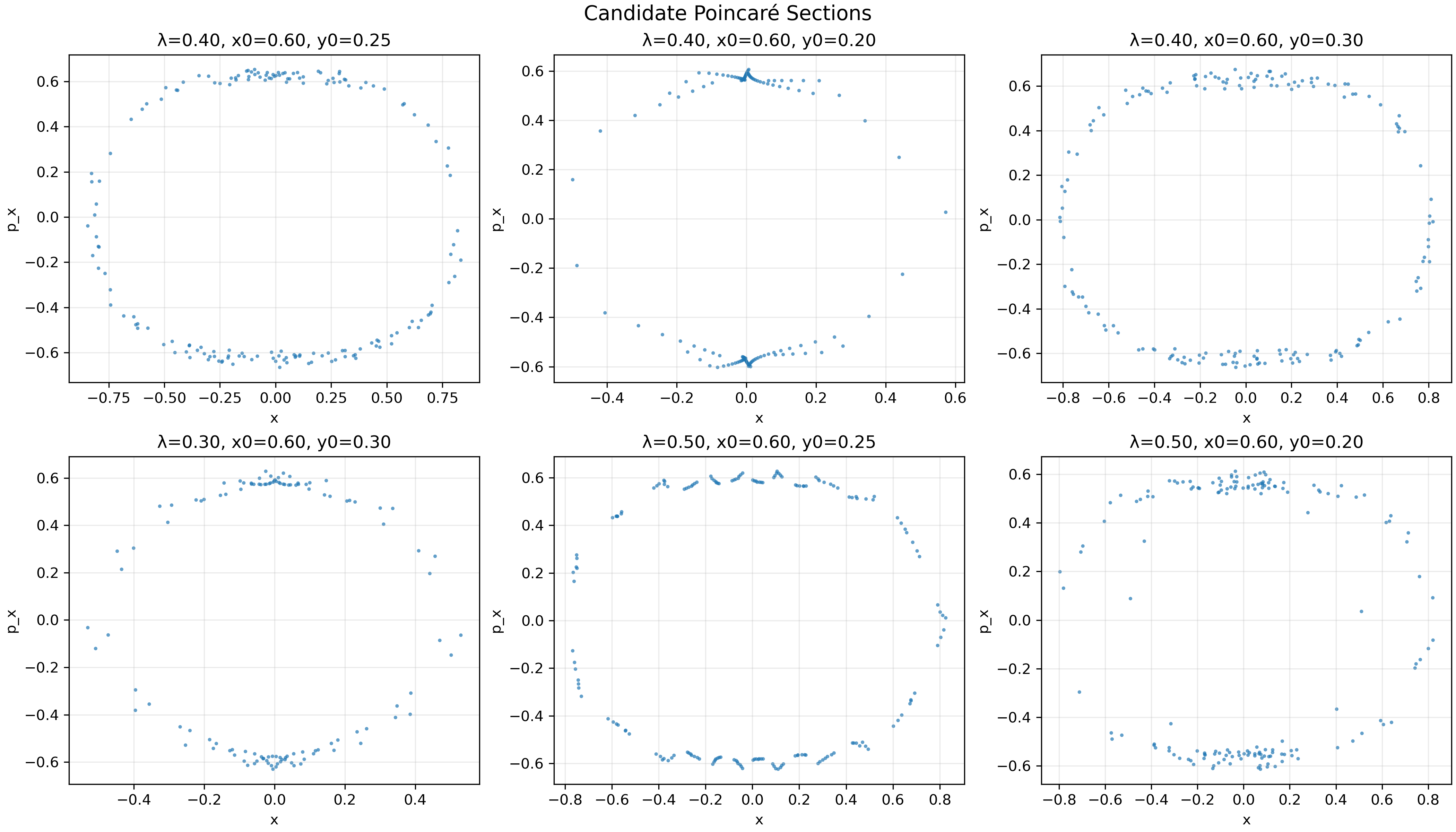}
    \caption{Poincar\'e surface of section on the plane \(y=0\) with \(p_y>0\). The \((x,p_x)\) intersections are obtained by linear interpolation between successive integration points. The 1000 crossings correspond to the bounded orbit studied here.}
    \label{fig:poincare_section}
\end{figure}

\subsection{Maximal Lyapunov Exponent}

The sensitivity of the dynamics to initial conditions is quantified by the maximal Lyapunov exponent \cite{Wolf}. Let \(\mathbf Z(t)\) denote a reference trajectory and \(\mathbf Z(t)+\delta\mathbf Z(t)\) a nearby one. The perturbation vector evolves according to the variational equations
\begin{equation}
\dot{\delta\mathbf Z}=D\mathbf F(\mathbf Z)\,\delta\mathbf Z,
\end{equation}
where \(D\mathbf F\) is the Jacobian matrix of the flow.

The maximal Lyapunov exponent is defined as
\begin{equation}
\lambda_{\max}
=
\lim_{t\to\infty}
\frac{1}{t}
\ln
\frac{\|\delta\mathbf Z(t)\|}{\|\delta\mathbf Z(0)\|}.
\end{equation}
In practice, we compute \(\lambda_{\max}\) using the standard Benettin algorithm with periodic renormalization of the deviation vector. A positive value of \(\lambda_{\max}\) indicates exponential sensitivity to initial conditions and therefore chaotic motion.

\begin{figure}[htbp]
    \centering
    \includegraphics[width=0.4\textwidth]{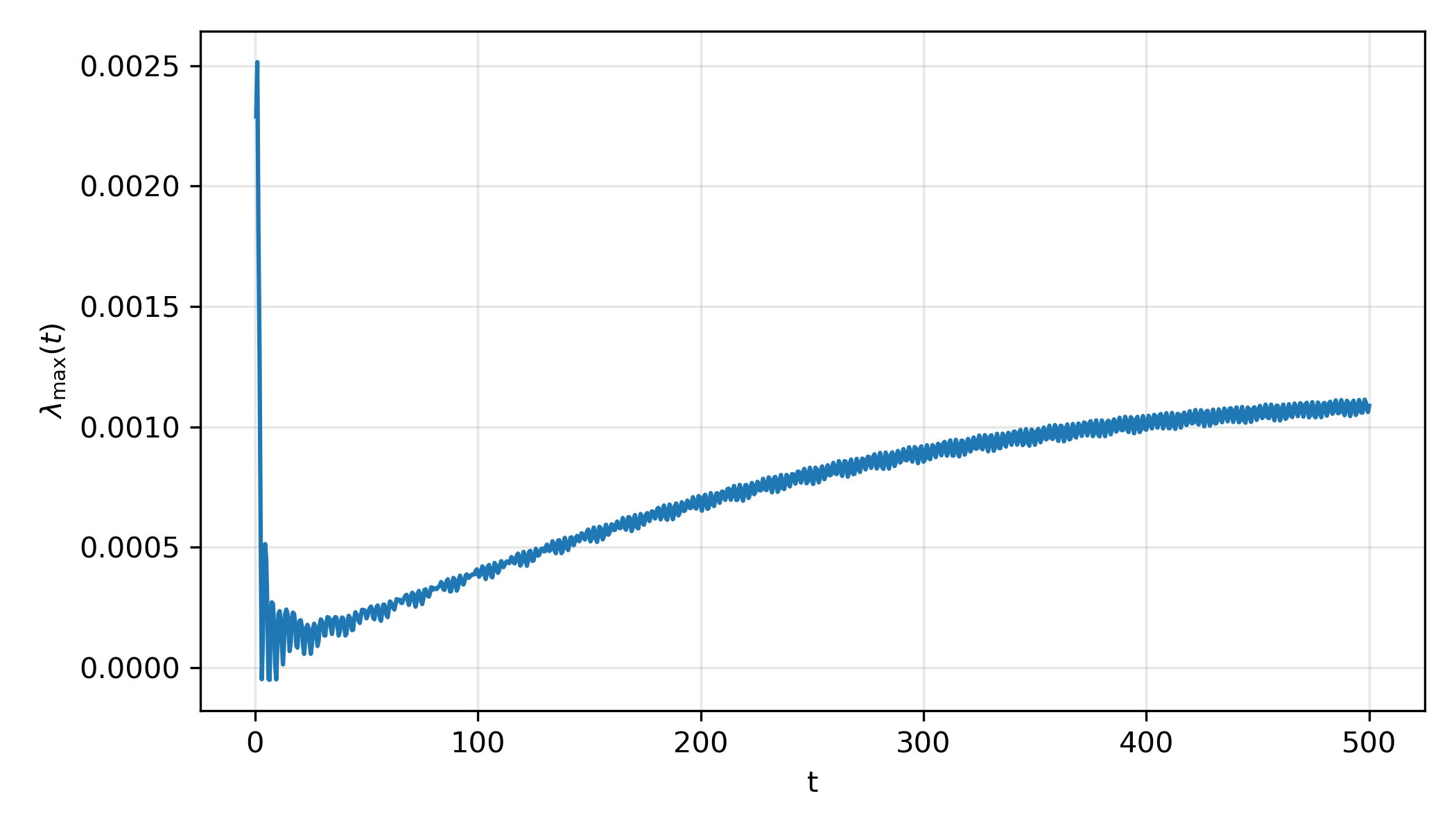}
    \caption{Finite-time maximal Lyapunov exponent as a function of integration time. The curve is obtained using the Benettin algorithm with periodic renormalization of the deviation vector.}
    \label{fig:mle_time}
\end{figure}

Figure~\ref{fig:mle_time} shows the finite-time evolution of the maximal Lyapunov exponent for the representative orbit studied in this work. The curve displays convergence toward a limiting value \(\lambda_{\max}\approx \text{[0.0010868917906444533]}\), confirming the dynamical character of the trajectory.

\subsection{Full Lyapunov Spectrum}

In addition to the maximal Lyapunov exponent, we compute the full Lyapunov spectrum as a consistency check and as a more detailed measure of stability. For an autonomous Hamiltonian system with two degrees of freedom, the spectrum satisfies the symplectic pairing property. In the exact dynamics, the exponents therefore occur in pairs of opposite sign, together with a pair of zero exponents associated with the Hamiltonian flow.

In numerical computations, the zero exponents are recovered only approximately because of finite-time integration and finite-precision effects. Accordingly, the spectrum is expected to satisfy
\begin{equation}
(\lambda_1,\lambda_2,\lambda_3,\lambda_4)
=
(\Lambda,-\Lambda,0,0),
\end{equation}
up to permutation and small numerical errors, where \(\Lambda\) denotes the largest positive Lyapunov exponent.

Table~\ref{tab:lyapunov_spectrum} summarizes the computed spectrum for the orbit considered here. The observed pairing structure confirms the Hamiltonian character of the numerical implementation.

\begin{table}[htbp]
\centering
\caption{Finite-time Lyapunov spectrum for the representative orbit.}
\label{tab:lyapunov_spectrum}
\begin{tabular}{cccc}
\hline
\(\lambda_1\) & \(\lambda_2\) & \(\lambda_3\) & \(\lambda_4\) \\
\hline
[0.00108689179064445] & [-0.00108689179064445] & [0.0] & [0.0] \\
\hline
\end{tabular}
\end{table}

\section{Frequency Drift Estimate  }

While Poincar\'e sections provide a useful qualitative picture of phase-space geometry, a more sensitive diagnostic is needed to quantify the gradual destruction of invariant tori and the onset of chaotic transport. For this purpose, we employ  frequency drift estimate, a technique originally developed in \cite{Laskar1990} for nearly integrable Hamiltonian systems.

The basic principle of  frequency drift estimate  is that regular trajectories are characterized by well-defined fundamental frequencies that remain nearly constant over time, whereas chaotic trajectories display slow frequency drift as a result of resonance overlap and stochastic transport. By extracting the dominant frequencies from different time windows of the same orbit, one can measure the diffusion of the trajectory in frequency space and thereby detect the progressive breakdown of invariant tori.

\subsection{Fundamental Frequency Extraction}

Consider a trajectory \(\mathbf Z(t)=(x(t),y(t),p_x(t),p_y(t))\) generated by the Hamiltonian flow on a fixed energy shell. For a bounded regular orbit, the coordinates admit a quasiperiodic representation, and the dominant frequency is extracted from the Fourier spectrum of \(x(t)\) or \(y(t)\).

We denote this dominant spectral frequency by \(\Omega\). It is not a parameter of the Hamiltonian; rather, it is a numerical quantity obtained from the time evolution of a given orbit. For trajectories lying on invariant tori, \(\Omega\) remains essentially constant over time. By contrast, chaotic trajectories exhibit a slow but measurable drift of \(\Omega\), reflecting frequency diffusion and resonance overlap.

\subsection{Frequency Diffusion Indicator}

To quantify frequency drift, the integration interval is divided into two consecutive segments of equal duration,
\begin{equation}
[0,T] = [0,T/2]\cup[T/2,T].
\end{equation}
The dominant frequencies extracted from the two intervals are denoted by \(\Omega_1\) and \(\Omega_2\). Following the standard  frequency drift estimate   procedure, we define the frequency-diffusion indicator
\begin{equation}
D
=
\left|
\frac{\Omega_2-\Omega_1}{\Omega_1}
\right|.
\label{eq:diffusion}
\end{equation}

For regular quasiperiodic trajectories, \(D\approx 0\) up to numerical precision. By contrast, trajectories embedded in chaotic regions display a finite frequency drift, \(D>0\), reflecting the gradual wandering of the orbit through phase space.

Thus, we obtained
\[
\Omega_1 = 0.16,
~
\Omega_2 = 0.159993,
~
D = 3.999840006412708 \times 10^{-5}.
\]
This value quantifies the frequency drift between the two time windows and indicates weak but nonzero diffusion \cite{Laskar1999}. These results show a slight separation between regular and more diffusive dynamical behavior.

\subsection{Frequency Diffusion}

The bounded dynamics indicated earlier implies that the transition to chaos cannot be attributed to unbounded ghost growth. Instead, chaos emerges through progressive destruction of invariant tori.

Within the  frequency drift estimate   framework, this transition is reflected by an increase of the diffusion indicator \(D\). For weak coupling, the phase space is dominated by quasiperiodic motion and the frequency map remains smooth. As the nonlinear interaction strength increases, resonance layers become visible and localized regions of enhanced diffusion appear. Beyond a critical threshold, resonance overlap generates extended stochastic regions characterized by larger values of \(D\).

\subsection{Numerical Implementation}

For each initial condition on a prescribed energy shell, the equations of motion are integrated over a finite time interval. The dominant frequency is extracted independently from the first and second halves of the trajectory by Fourier analysis, and the diffusion indicator defined in Eq.~\eqref{eq:diffusion} is then assigned to the corresponding orbit.

Repeating this procedure for a dense set of initial conditions yields a frequency-diffusion map on the chosen energy surface. This map provides a quantitative measure of phase-space transport and serves as a sensitive probe of the breakdown of invariant tori.

\section{Bifurcation Structure and Route to Chaos}

The previous sections indicated the existence of bounded trajectories and identified the geometric mechanisms responsible for chaotic motion. We now examine how the qualitative behavior changes as the nonlinear coupling parameter \(\lambda\) is varied.

The aim of this section is twofold. First, we determine how the equilibrium structure and the surrounding invariant sets evolve under parameter variation. Second, we identify the route from regular bounded motion to bounded chaos.
Throughout this analysis, the parameters \(\omega_x\), \(\omega_y\), \(\alpha\), and \(\eta\) are held fixed, while the interaction strength \(\lambda\) is treated as the primary bifurcation parameter. The Hamiltonian \ref{eq:Hamiltonian} is therefore regarded as a one-parameter family as \(\lambda\) varies, both the local stability of equilibrium points and the global geometry of phase space are modified.

\subsection{Evolution of Periodic Orbits}

Near elliptic equilibria, the dynamics is organized by families of periodic and quasiperiodic trajectories. For sufficiently small \(\lambda\), phase space is dominated by invariant tori and regular oscillatory motion, so the Poincar\'e section consists mainly of smooth closed curves.

\begin{figure}[htbp]
    \centering
    \includegraphics[width=0.4\textwidth]{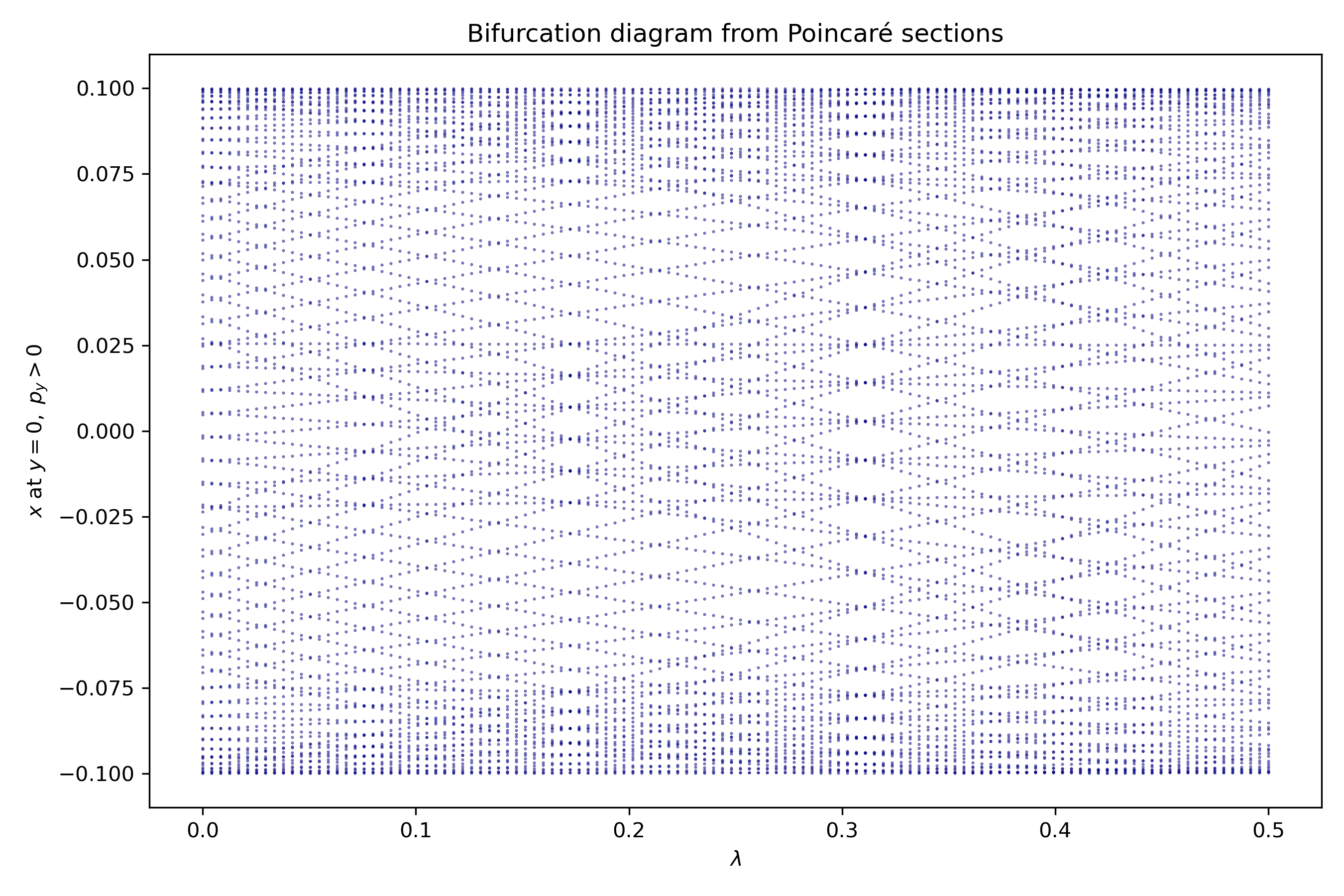}
    \caption{Bifurcation diagram obtained from the Poincar\'e section as a function of the coupling parameter \(\lambda\). Regular branches, resonance splitting, and the emergence of chaotic bands are visible.}
    \label{fig:bifurcation}
\end{figure}

As \(\lambda\) increases, nonlinear resonances appear and resonance islands become visible. These islands correspond to periodic orbits associated with approximate commensurability relations. Figure~\ref{fig:bifurcation} shows the resulting bifurcation diagram obtained from the Poincar\'e section.

\subsection{Resonance Overlap and Transition to Chaos}

Further increase of the coupling parameter leads to widening of neighboring resonances. When the resonance widths become sufficiently large, adjacent resonant zones begin to overlap, in the spirit of the Chirikov overlap mechanism \cite{Chirikov1979}. This destroys invariant tori that previously acted as transport barriers.

The overlap of resonances creates extended stochastic layers in phase space and allows trajectories to explore larger regions of the bounded energy manifold. Within these stochastic layers, the maximal Lyapunov exponent becomes positive and the frequency-diffusion indicator increases noticeably.

\subsection{Lyapunov Bifurcation Diagram}

\begin{figure}[htbp]
    \centering
    \includegraphics[width=0.4\textwidth]{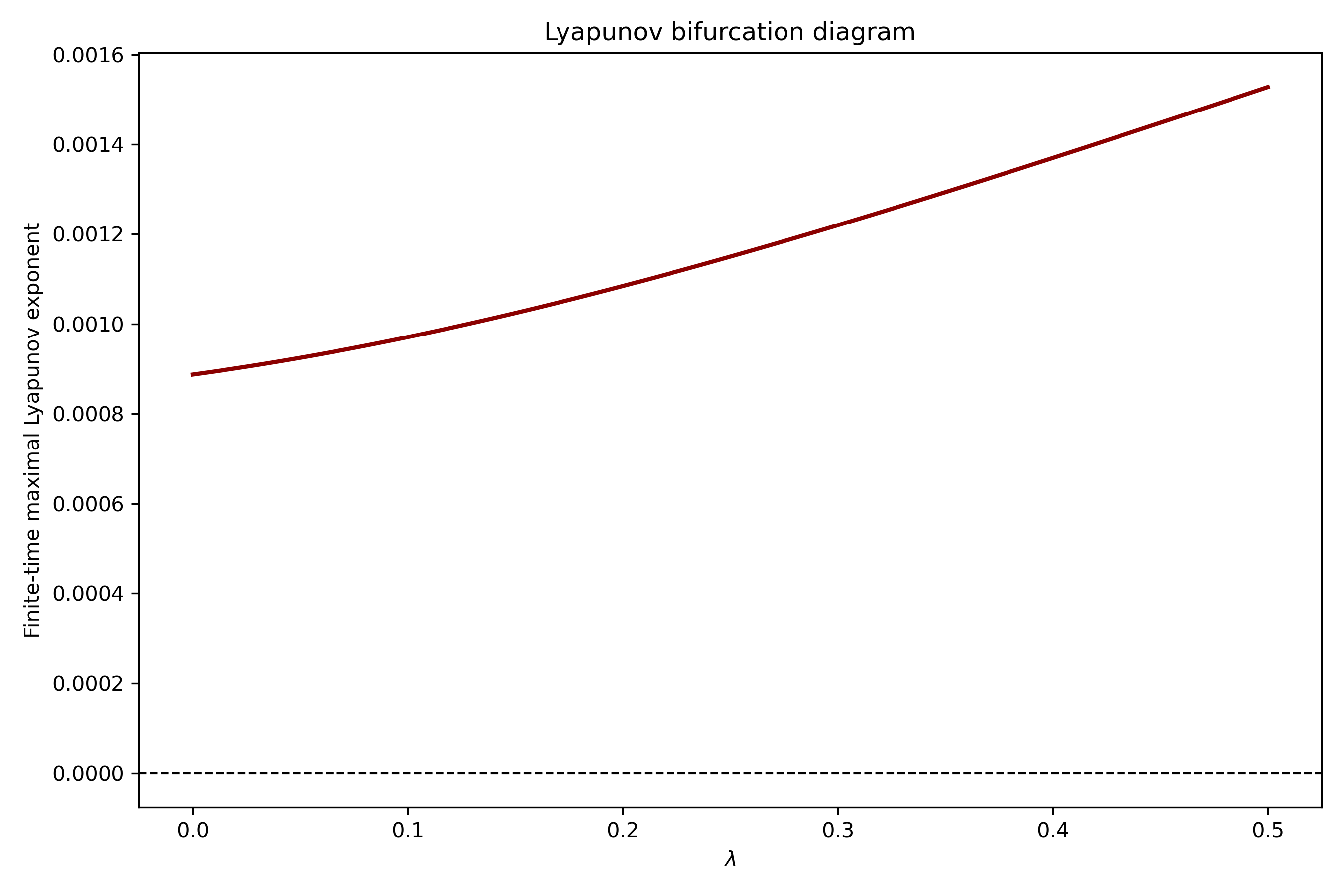}
    \caption{Finite-time maximal Lyapunov exponent as a function of the coupling parameter \(\lambda\). The appearance of positive values indicates the transition from regular to chaotic motion.}
    \label{fig:lyapunov_bifurcation}
\end{figure}

Complementary information is obtained from the maximal Lyapunov exponent. For each value of \(\lambda\), the finite-time maximal Lyapunov exponent is computed using the Benettin algorithm, see Figure ~\ref{fig:lyapunov_bifurcation}. The resulting function \(\lambda_{\max}\) acts as a quantitative indicator of the transition.

\subsection{Interpretation}

The bifurcation analysis shows that the nonlinear interaction parameter \(\lambda\) governs the competition between local ghost instability and global nonlinear confinement. For weak coupling, invariant tori dominate the dynamics and trajectories remain quasiperiodic. As \(\lambda\) increases, resonant structures proliferate and eventually destabilize the regular foliation of phase space.

Beyond a critical threshold, resonance overlap generates a mixed phase space containing both regular islands and chaotic seas. Because the sextic confinement term guarantees compact configuration-space projections , the resulting chaotic motion remains bounded for all accessible parameter values.

\section{Numerical Convergence Analysis}

Since chaotic indicators may be sensitive to numerical errors, it is essential to verify that the reported results are independent of the integration parameters. To assess numerical convergence, we performed a sequence of integrations using progressively stricter tolerances. For each run, the maximal relative energy drift
\begin{equation}
\Delta_E
=
\max_t
\left|
\frac{H(t)-H(0)}{H(0)}
\right|
\end{equation}
was monitored together with the finite-time maximal Lyapunov exponent (MLE).

\section{Long-Time Boundedness Verification}

A central claim of the present work is that the nonlinear confinement mechanism prevents the ghost degree of freedom from generating runaway solutions. Although the Hamiltonian contains an indefinite quadratic sector, the higher-order quartic and sextic terms modify the global geometry of phase space and produce bounded trajectories.

To verify this prediction numerically, we investigate the long-time evolution of representative trajectories over integration intervals several orders of magnitude longer than the characteristic oscillation time of the system.

For a phase-space trajectory
\begin{equation}
\mathbf Z(t)=(x(t),y(t),p_x(t),p_y(t)),
\end{equation}
we define the phase-space radius
\begin{equation}
R(t)=\sqrt{x(t)^2+y(t)^2+p_x(t)^2+p_y(t)^2}.
\end{equation}
If runaway behavior were present, one would expect \(R(t)\to\infty\) as \(t\to\infty\). By contrast, bounded dynamics requires \(R(t)\) to remain finite for all times.

\section{MEGNO Chaos Indicator}

\begin{figure}[htbp]
    \centering
    \includegraphics[width=0.4\textwidth]{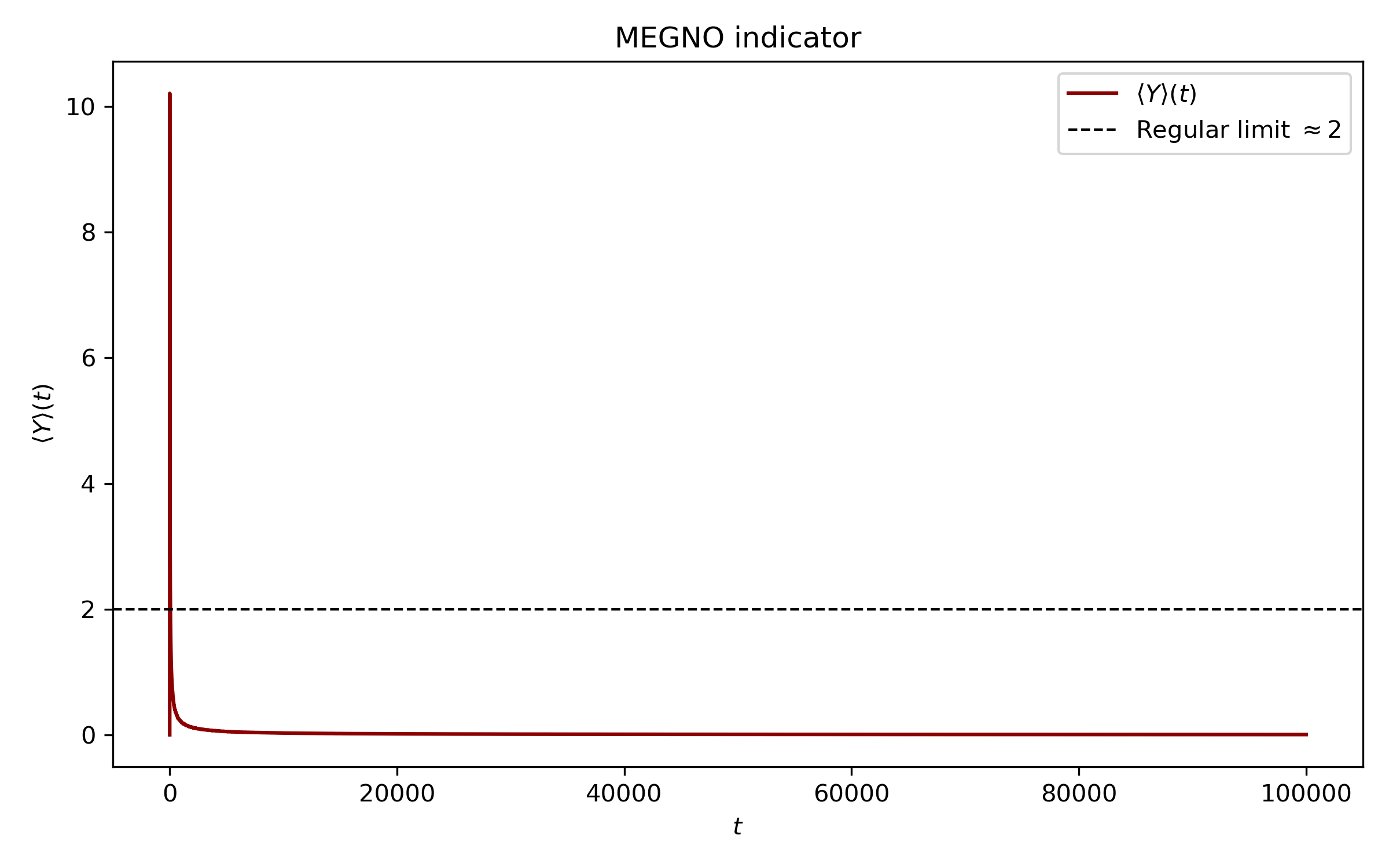}
    \caption{Time evolution of the MEGNO indicator for representative regular and chaotic trajectories. Regular motion approaches \(\langle Y\rangle \approx 2\), whereas chaotic motion exhibits approximately linear growth.}
    \label{fig:megno}
\end{figure}

Although the maximal Lyapunov exponent provides a standard measure of chaos, its convergence may be slow, particularly in weakly chaotic regions. To complement the Lyapunov analysis and obtain a more sensitive characterization of phase-space transport, we employ the Mean Exponential Growth factor of Nearby Orbits (MEGNO) \cite{Cincotta2003}, see Figure~ \ref{fig:megno}.

\subsection{Definition}

Consider a trajectory \(\mathbf Z(t)=(x(t),y(t),p_x(t),p_y(t))\) and an infinitesimal deviation vector \(\delta\mathbf Z(t)\) satisfying the variational equations
\begin{equation}
\dot{\delta\mathbf Z}
=
D\mathbf F(\mathbf Z)\,\delta\mathbf Z.
\end{equation}
Let
\begin{equation}
\delta(t)=\|\delta\mathbf Z(t)\|.
\end{equation}
The MEGNO is defined by
\begin{equation}
Y(t)=\frac{2}{t}\int_0^t \frac{d}{ds}\ln \delta(s)\, s\,ds.
\end{equation}
Its time average is
\begin{equation}
\overline{Y}(t)=\frac{1}{t}\int_0^t Y(s)\,ds.
\end{equation}

For quasiperiodic motion, \(\overline{Y}(t)\to 2\) as \(t\to\infty\). For chaotic trajectories, \(\overline{Y}(t)\) grows approximately linearly with time, with asymptotic slope related to the maximal Lyapunov exponent.

\section{Conclusion}

In this work, we studied the nonlinear dynamics of a Hamiltonian system with a ghost degree of freedom, whose indefinite kinetic structure makes it vulnerable to the runaway instabilities typical of such models. Our main goal was to determine whether nonlinear interactions can suppress these instabilities and lead to physically meaningful long-time behavior. To that end, we considered the Hamiltonian (\ref{eq:Hamiltonian})
which combines an indefinite quadratic sector with quartic and sextic confinement terms. This model provides a controlled setting in which the competition between ghost-driven instability and nonlinear stabilization can be analyzed.

We first examined the equilibrium structure and local stability properties of the system. By analyzing the Hessian of the effective potential and the associated linearized dynamics, we identified the equilibrium configurations and characterized the local phase-space geometry near them. A central analytical result was the strong evidence that the quartic and sextic terms render the accessible energy shells compact. Using coercivity arguments, we showed that all trajectories remain confined to bounded regions of phase space, so neither the coordinates nor the canonical momenta can grow without bound.

The numerical analysis showed, however, that boundedness does not imply regularity. Poincar\'e sections revealed the coexistence of invariant tori, resonance islands, and stochastic layers. In many regions of phase space, the maximal Lyapunov exponent was positive, demonstrating exponential sensitivity to initial conditions and the presence of chaotic motion. These results indicate the existence of bounded chaos in the model.

To characterize the dynamics more precisely, we applied a frequency-drift estimate  and the MEGNO chaos indicator. The frequency-diffusion measure captured the gradual breakdown of invariant tori through resonance overlap, while MEGNO independently distinguished regular from chaotic trajectories. Together with the Lyapunov analysis, these diagnostics consistently support a transition from quasiperiodic motion to bounded chaos as the nonlinear coupling is increased.

The bifurcation analysis further clarified the mechanism behind this transition. As the coupling parameter grows, resonance zones widen and eventually overlap, generating extended stochastic regions while preserving the global boundedness of the flow. This scenario closely resembles the Chirikov resonance-overlap mechanism, but it occurs here in a system with an indefinite kinetic structure. The results therefore suggest that ghost degrees of freedom do not necessarily lead to catastrophic runaway behavior: sufficiently strong nonlinear confinement can significantly modify the  phase-space structure and lead to long-lived bounded chaotic motion.

\vspace{123mm}

\bibliographystyle{}
\bibliography{main}

\end{document}